\documentclass{article}

\usepackage{arxiv}
\usepackage{eurosym}
\usepackage[utf8]{inputenc} 
\usepackage[T1]{fontenc}    
\usepackage{hyperref}       
\usepackage{url}            
\usepackage{booktabs}       
\usepackage{amsfonts}       
\usepackage{nicefrac}       
\usepackage{microtype}      
\usepackage{lipsum}
\usepackage{graphicx}
\usepackage{color}
\usepackage{amssymb}
\usepackage[ruled,vlined]{algorithm2e}
\usepackage{subcaption}
\usepackage{amsmath}
\usepackage[autostyle]{csquotes}

\title{Improving in-home appliance identification using fuzzy-neighbors-preserving analysis based QR-decomposition}

\author{
  Yassine Himeur\thanks{This paper has been accepted in Fifth International Congress on Information and Communication Technology (ICICT), London, UK, 2020.} , Abdullah Alsalemi, Faycal Bensaali\\
  Department of Electrical Engineering\\
  Qatar University\\
  Doha, Qatar \\
  \texttt{yassine.himeur@qu.edu.qa;a.alsalemi@qu.edu.qa;f.bensaali@qu.edu.qa} \\
   \And
 Abbes Amira \\
  Institute of Artificial Intelligence\\
  De Montfort University\\
  Leicester, United Kingdom \\
  \texttt{abbes.amira@dmu.ac.uk} \\
}

\begin{document}
\maketitle

\begin{abstract}
This paper proposes a new appliance identification scheme by introducing a novel approach for extracting highly discriminative characteristic sets that can considerably distinguish between various appliance footprints. In this context, a precise and powerful characteristic projection technique depending on fuzzy-neighbors-preserving analysis based QR-decomposition (FNPA-QR) is applied on the extracted energy consumption time-domain features. The FNPA-QR aims to diminish the distance among the between class features and increase the gap among features of dissimilar categories. Following, a novel bagging decision tree (BDT) classifier is also designed to further improve the classification accuracy. The proposed technique is then validated on three appliance energy consumption datasets, which are collected at both low and high frequency. The practical results obtained point out the outstanding classification rate of the time-domain based FNPA-QR and BDT. 
\end{abstract}


\section{Introduction} \label{sec1}
Various studies claim that domestic buildings are responsible of up to 40\% of the total energy use around the globe in the past decades \cite{RASHID2019796}. A large part of this energy is wasted because of inappropriate usage habits and absence of data highlighting the consumption rate of each device in residential area. In addition, individuals have not fully harnessed possibilities to reduce their wasted energy. Consequently, creating an energy efficiency lifestype begins with i) providing them with fine-grained power use profiles at real-time, ii) helping them comprehend their consumption behavior and iii) encouraging them to make energy saving actions through allowing tailored recommendations \cite{Alsalemi2018IEESyst}. Therefore, it would be of great positive potential if smartphones can alert users with notifications about power consumption (PC) level of each appliance and provides appliance usage status \cite{Alsalemi2019SBC,Sardinos2019}. Currently, sub-meters and smart sensors are the candidate solutions to collect this kind of data. The first one provides aggregated consumption profiles without permitting to pick up individual loads, while the second category is very costly to be deployed in residential buildings \cite{Li2019SG,ALSALEMI2019,Alsalem2019GCGW}.   

In that direction, energy disaggregation for appliance identification becomes increasingly a hot area of research. Individual consumption footprints can be picked up by sampling the whole PC signal through making use of a non-intrusive load monitoring (NILM) scheme \cite{Welikala2019SG}. The proposed framework focuses on the design of an efficient appliance identification system using a robust feature extraction scheme based on dimensionality reduction. First, time-domain (TD) descriptors are implemented to capture relevant features from PC footprints. They can pick up step-changes in PC signals generally occurring when an appliance is switched on/off. Second, a fuzzy-neighbors-preserving analysis based QR-decomposition, namely FNPA-QR, is introduced to reduce the dimension of TD signatures. In this context, as the PC signals have an important variance nature, the extracted TD data from the PC signals are usually scattered freely within the initial feature ensemble. Third, a bagging decision tree (BDT) classifier is designed to robustly identify each appliance category. The BDT exploits multiple weak machine learning models to develop a nearly optimal classifier that outperform other classification models, especially when it is applied on TD features.  

The remaining of this paper consists of the following parts. The proposed appliance identification system is explained in Section \ref{sec2}. Section \ref{sec3} presents the obtained empirical results with a set of comparisons at different levels of the proposed appliance recognition architecture. Finally, Section \ref{sec4} summarizes the main outcomes of this framework. 

\section{Proposed system} \label{sec2}
\begin{figure*}[t!]
\begin{center}
\includegraphics[width=14.8cm, height=7.2cm]{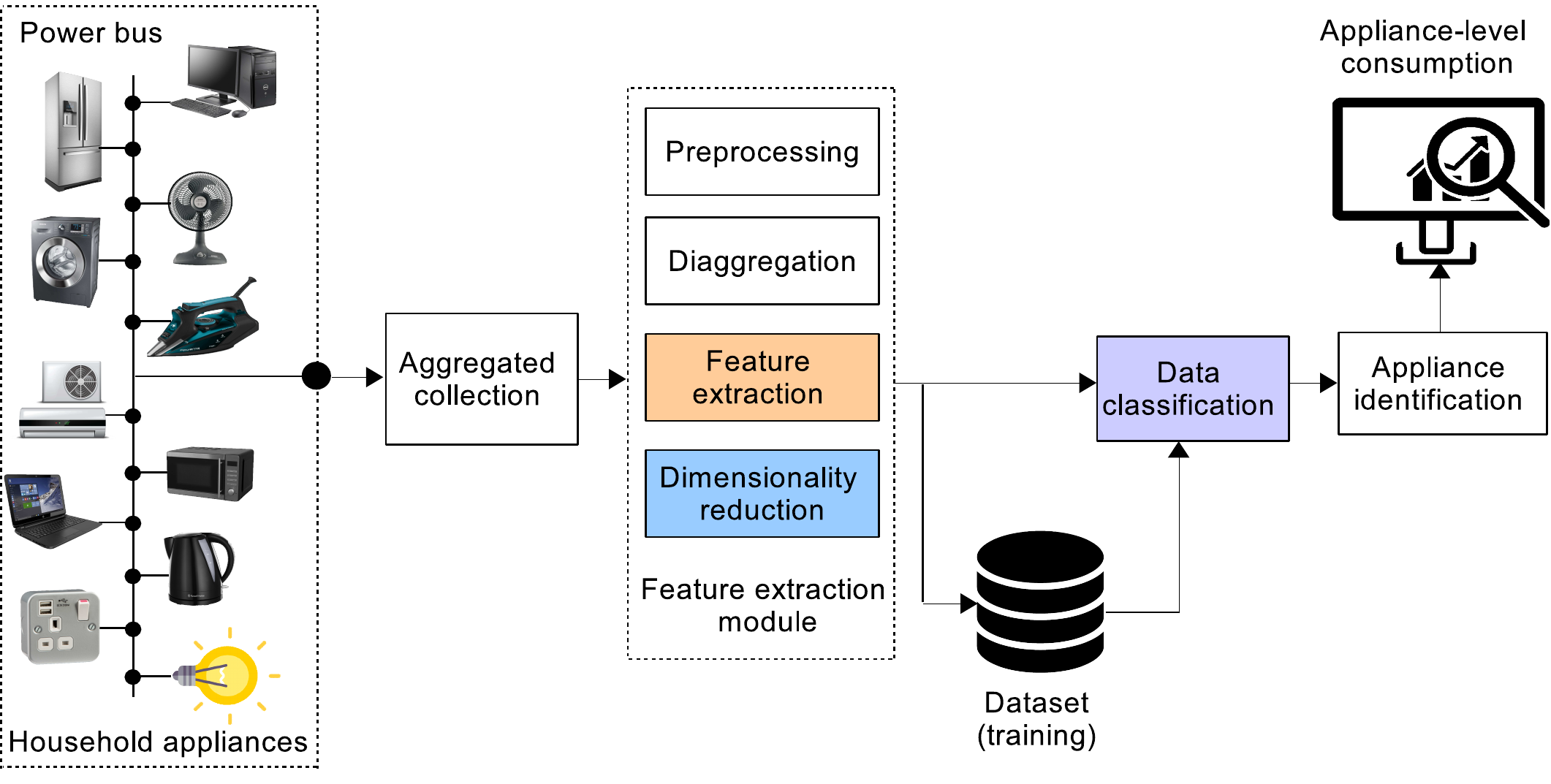}
\end{center}
\caption{Block diagram of the proposed in-home appliance identification system.}
\label{FlowChart}
\end{figure*}

In this section, contributions of this framework are highlighted through explaining in details the main modules proposed to design a robust in-home appliance recognition architecture. Fig. \ref{FlowChart} portrays the block-diagram of the proposed system.

\subsection{Feature extraction using TD descriptors}
In order to select the best TD descriptor for the appliance recognition purpose, five TD descriptors are firstly considered and their performance are compared. The best descriptor is then implemented to design the appliance identification system. In order to describe the TD feature descriptors used under our framework, let us consider a sampled version of the energy consumption signal denominated as: $y[i]$, with $i = 1,2, \cdots, M$, of length $M$ and collected at the sampling frequency $f_{y}$, in order to extract the TD features, a windowing process is applied on $y$ where a window length $N$ is employed and the TD property $Y(k)$ of each window $k$ $(k=1,2, \cdots, K)$ is then collected as follows:   

\noindent - Root mean square feature (RMSF)
\begin{equation}
Y_{RMS}(k)=\textstyle\sum\limits_{i=1}^{N}\sqrt{\frac{1}{N}(y_{i}^{2})}
\end{equation}

\noindent - Mean absolute deviation feature (MADF)
\begin{equation}
Y_{MAD}(k)=\textstyle\sum\limits_{i=1}^{N}\frac{1}{N}\left\vert y_{i}-\mu \right\vert 
\end{equation}
where $\mu$ represent the central tendency,

\noindent - Integrated absolute magnitude feature (IAMF)
\begin{equation}
Y_{IAMF}(k)=\frac{1}{N}\textstyle\sum\limits_{i=1}^{N}\frac{y_{i}^{2}}{2}\textnormal{sgn}%
(y_{i})+\mu 
\end{equation}

\noindent - Waveform length feature (WLF)
\begin{equation}
Y_{WLF}(k)=\log \left( \textstyle\sum\limits_{i=1}^{N-1}\left\vert
y_{i+1}-y_{i}\right\vert \right) =\log \left(
\textstyle\sum\limits_{i=1}^{N-1}\left\vert \Delta y_{i}\right\vert \right) 
\end{equation}

\noindent - Slope sign change feature (SSCF)
\begin{equation}
S_{SSC}(k)=\textstyle\sum\limits_{i=2}^{N-1}f[(y_{i}-y_{i-1})\times (y_{i}-y_{i+1})] 
\end{equation}
where
\begin{equation}
f(y)=\left\{ 
\begin{array}{cc}
1 & \textnormal{if }y\geq \textnormal{threshold} \\ 
0 & \textnormal{otherwise \ \ \ \ \ \ \ \ }%
\end{array}%
\right. 
\end{equation}


\subsection{Dimensionality reduction using FNPA-QR}
The FNPA-QR is applied on the TD signals, which is a variant of fuzzy-linear discriminant analysis (FLDA). The latter has been used to study the class relationship between samples, however the main drawback of FLDA is related to the fact that it could not find out the regional geometric structure of samples. Accordingly, in discriminant analysis, the regional arrangement is more prominent in comparison to the global one \cite{Cai2007}. In addition, the discrimination ability between samples pertaining to different groups can be improved if the local structure is preserved.  To that end, the FNPA-QR introduces a novel feature projection scheme that can map the samples into a new subspace by analyzing the adjacent patterns. Consequently, it makes adjacent samples with the same label more close and in contrast, turns the adjacent coefficients with different labels to be far away. The main steps to perform the FNPA-QR approach are summarized.

\noindent 1. Set the features data using a matrix structure $Y(i,j)$, where $i$ is the index of feature vectors and $j$ is the number of patterns in each vector and set the number of reduced samples $r$

 \noindent 2. Estimate the within-class-scatter (WCS) array $Y_{W}$ as:
\begin{equation}
Y_{W}=\left( YDY-YWY^{T}\right) =YL_{1}Y^{T}  \label{eq12}
\end{equation} 
where $L_{1} = D - W$ represents the Laplacian array as described in \cite{He1388260,KHUSHABA2013}. $D$ constitutes a diagonal array that its coefficients are derived by summing the WCS array $W$. $W$ is the array of the WCS patterns; 
 
\noindent 3. Estimate the between-class-scatter (BCS) array $Y_{B}$ as:
\begin{equation}
Y_{B}=(MEM^{T}-MBM^{T})=ML_{2}M^{T}  \label{eq14}
\end{equation}%
where $M$ is the mean matrix of total patterns. $L_{2}= E - B$ and $E$ represents a diagonal array; its inputs are column sums of $B$. $B$ is the array of the BCS patterns;

\noindent 4. Estimate the transformation matrix $\mathbf{H}_{FNPA-QR}$ as follows
\begin{equation}
\mathbf{H}_{FNPA-QR}=\arg \textnormal{max}trace\left( \frac{\mathbf{H}^{T}Y_{B}%
\mathbf{H}}{\mathbf{H}^{T}Y_{W}\mathbf{H}}\right)  \label{eq15}
\end{equation}\
 
\noindent 5. Calculate the matrix $Q$ using $H = Q \times R$, where $R$ represents an upper-triangular array and $Q$ defines an orthogonal array; 

\noindent 6. Make $\mathbf{H}_{FNPA-QR} = Q$ and project the feature matrix with the transformation matrix as follows:
\begin{equation}
Y^{\prime }(i,r)=Y(i,r)\times H_{FNPA-QR}
\end{equation}

\subsection{Bagging decision tree (BDT) classifier}
\begin{figure}[t!]
\begin{center}
\includegraphics[width=15.5cm, height=6.2cm]{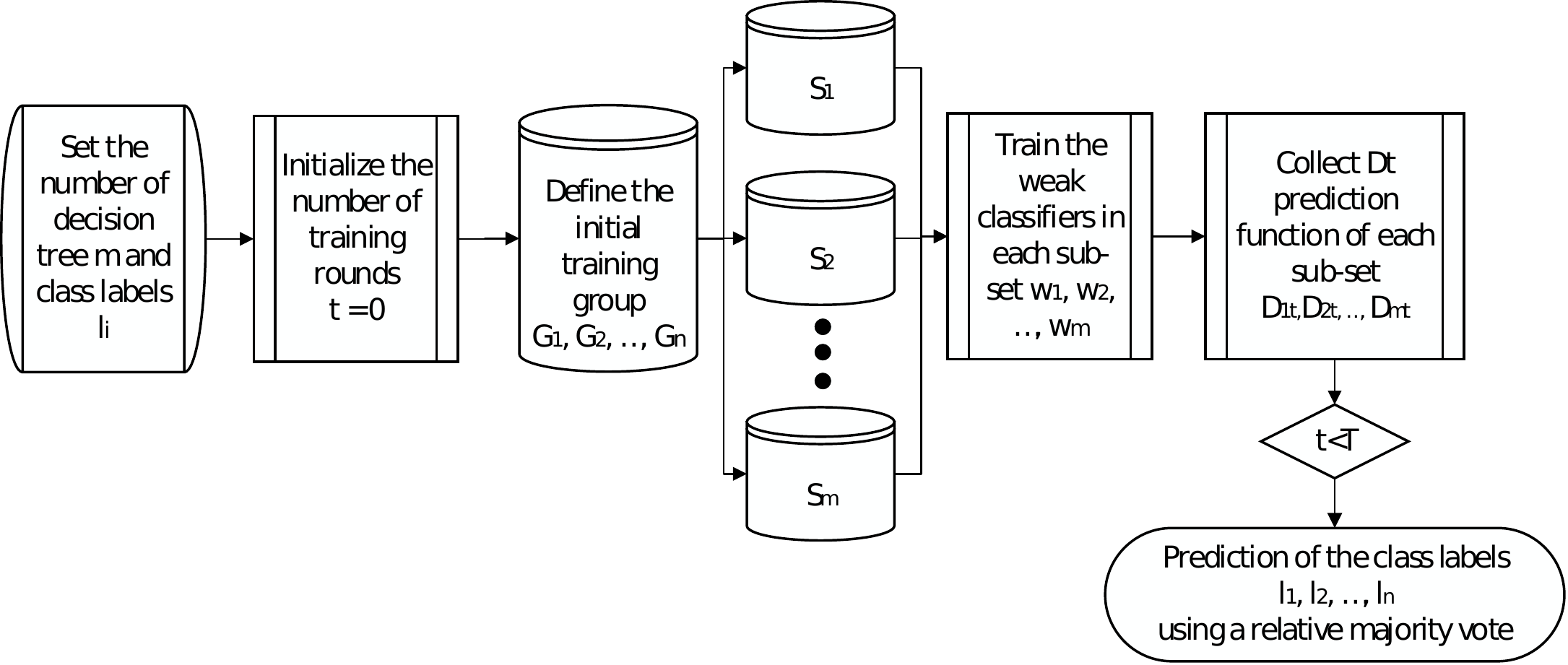}
\end{center}
\caption{Flowchart of a the BDT algorithm.}
\label{BDT}
\end{figure}

The idea behind BDT algorithm relies on the fact that the bagging process can trigger unstable weak classifiers to generate nearly optimal classifier. Instead of training the weak classifier on the global data set, each one is trained on a bootstrap set that is derived from the ensemble set. This makes the samples distribution along the training similar to the initial distribution. Consequently, the individual classifiers in a bagging ensemble can accurately classify the samples. Finally, a majority vote process is performed to normalize the performance of the BDT model and to additionally improve the classification accuracy. Fig. \ref{BDT} explains in details the steps required to implement the BDT classifier.

\section{Experimental results} \label{sec3}
The performance of the developed appliance in-home identification architecture is evaluated using three different energy consumption datasets. The ACS-f2 is the second version of the appliance consumption signature repository. It encompasses the electricity consumption footprints  of various electrical device categories using 0.1 Hz sampling frequency \cite{RIDI2014}. The GREEND: collects daily PC data of more than 35 domestic appliances deployed in 8 different households for a period ranging from 6-12 months. Additionally, consumption fingerprints are gathered using 1 Hz sampling frequency \cite{GREEND2014}. The WITHED: captures electricity consumption signatures for up to 47 appliance classes for a short period of 5 sec and using a 44000 Hz sampling rate. Under this framework, 11 appliance classes are employed to validate the proposed technique \cite{WHITED2016}.

\begin{figure}[t!]
\begin{center}
\includegraphics[width=7.5cm, height=5.3cm]{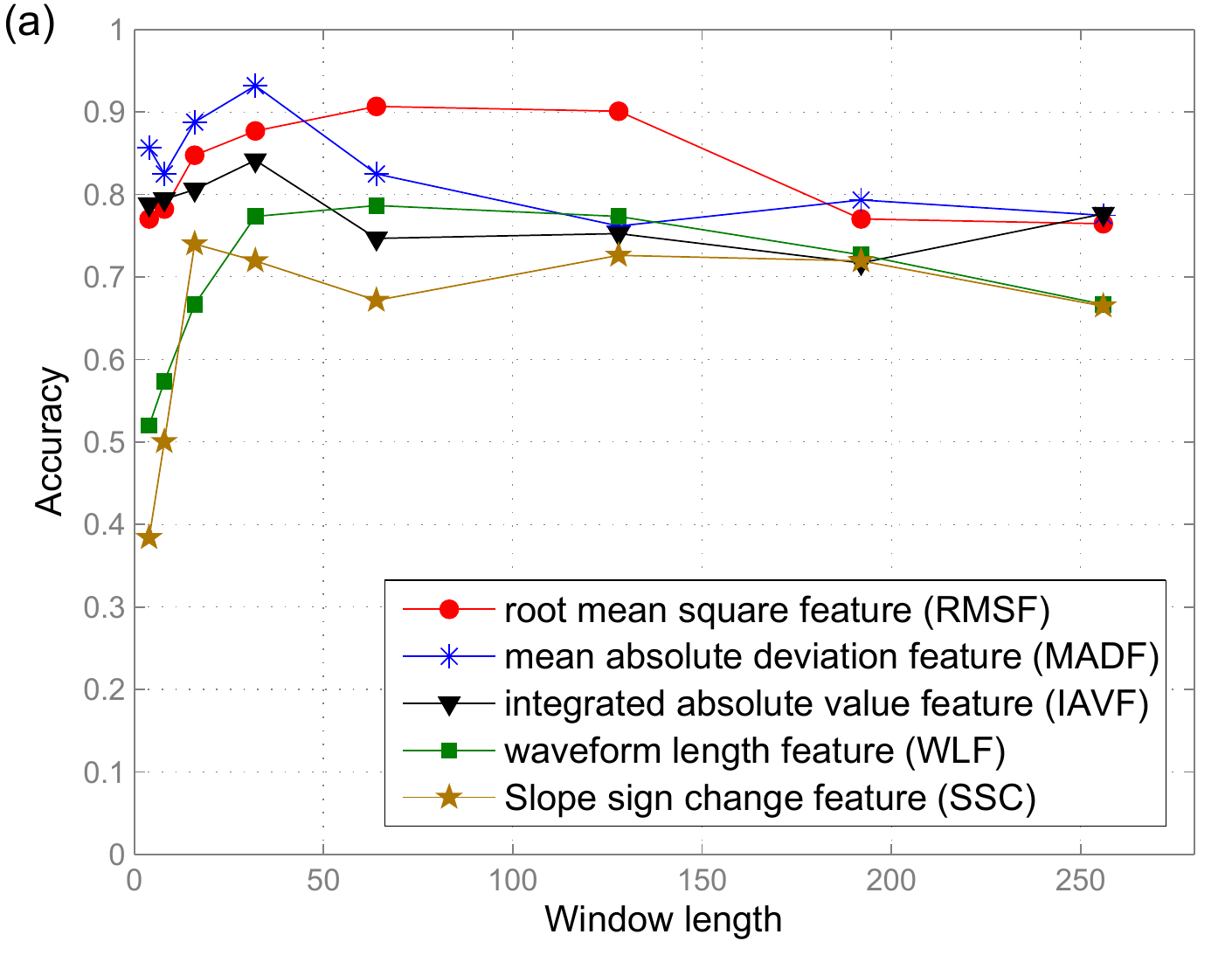}
\includegraphics[width=7.5cm, height=5.3cm]{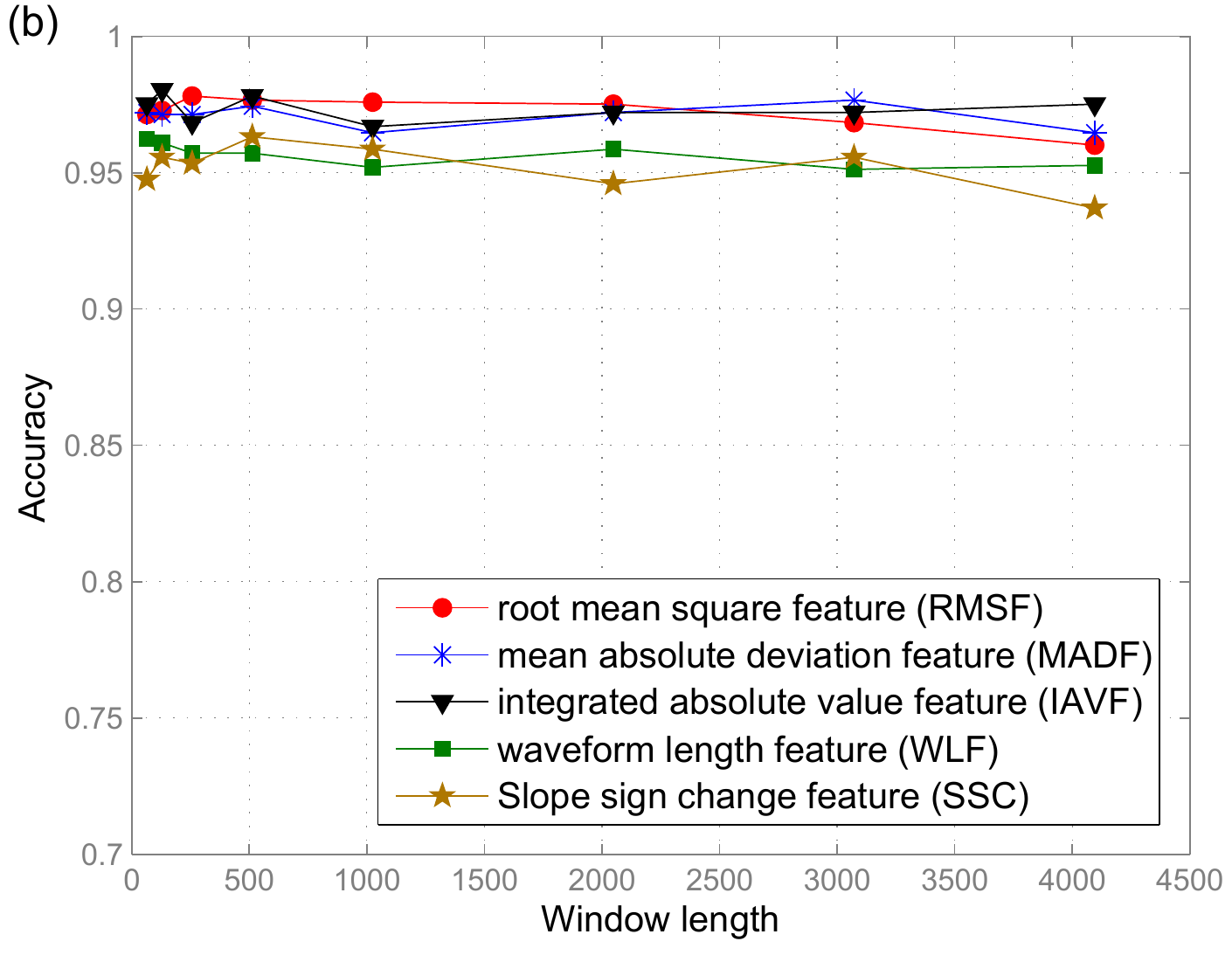}
\includegraphics[width=7.5cm, height=5.3cm]{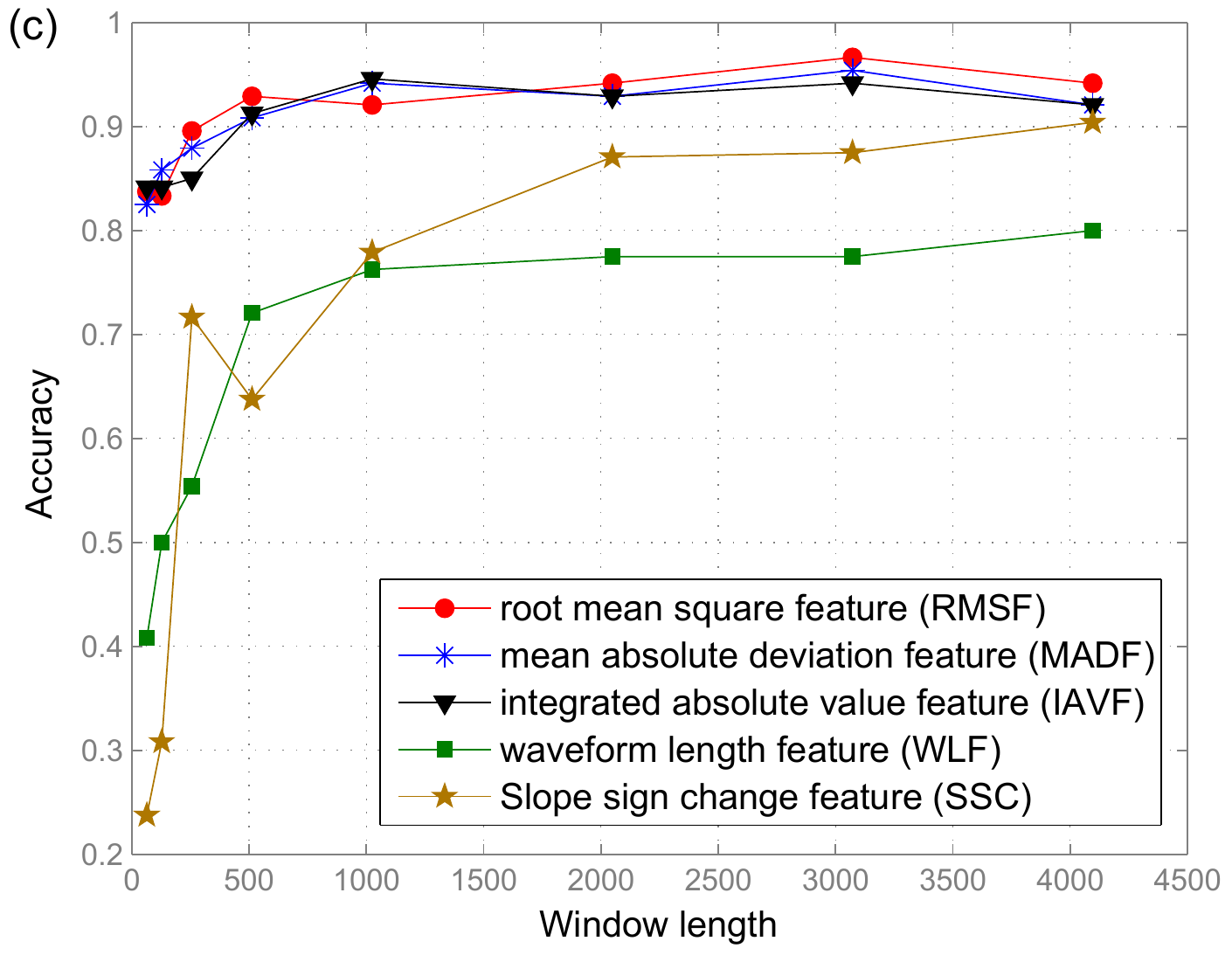}
\end{center}
\caption{Accuracy performance of different TD descriptors for the a) ACS-F2, b) GREEND and c) WHITED datasets.}
\label{ACC}
\end{figure}

First of all, the performance of the different TD descriptors using different window lengths is investigated with regard to the BDT classifier. The window lengths considered in the evaluation are selected according to the length of power consumption signals in each database. The accuracy outputs are then plotted in Fig. \ref{ACC}. It is clearly shown that RMSF descriptor can slightly outperform other descriptors, especially under window lengths of 2048 and 3072 for both GREEND and WHITED, and under a window length of 128 for ACS-F2. Furthermore, is has better performance stability than the other descriptors. Therefore, in the rest of this paper, the results of the proposed scheme are collected with reference to the RMSF descriptor using window lengths of 128 for ACS-F2 and 2048 for both GREEND and WHITED.

\subsection{Comparison versus other dimensionality reduction}
The performances the FNPA-QR system are validated with reference to other feature projection schemes, including principal component analysis (PCA), linear discriminant analysis (LDA) and FLDA for appliance identification. Table \ref{DimenComp} depicts the accuracy, F-score and computational complexity of the proposed technique based on FNPA-QR in comparison to other dimensionality reduction approaches. The outputs are collected using MATLAB 9.4 (R2018a) executed on a desktop that has a Core i7-3770S processor, 16 GRAM and 3.1 GHz. It is witnessed that the accuracy and F-score of the FNPA-QR are highly improved in comparison to other feature projection schemes. However, the time complexity is increased and this is due to the fact that FNPA-QR uses a fuzzy process along with the QR decomposition. 
\begin{table}[t!]
\caption{Performance comparison of FNPA-QR versus other dimensionality reduction techniques}
\label{DimenComp}
\begin{center}

\begin{tabular}{l|l|cccc}
\hline
{\small Dataset} & {\small Performance} & {\small PCA} & {\small LFDA} & 
{\small FLDA} & {\small FNPA-QR} \\ \hline
& {\small accuracy} & {\small 0.78} & {\small 0.83} & {\small 0.92} & 
{\small \textbf{0.99}} \\ 
{\small ACS-F2} & {\small F-score} & {\small 0.76} & {\small 0.82} & {\small %
0.92} & {\small \textbf{0.98}} \\ 
& {\small Time complexity (sec)} & {\small \textbf{0.035}} & {\small 1.81} & {\small %
0.71} & {\small 0.85} \\ \hline
& {\small accuracy} & {\small 0.86} & {\small 0.92} & {\small 0.95} & 
{\small \textbf{1}} \\ 
{\small GREEND} & {\small F-score} & {\small 0.85} & {\small 0.92} & {\small %
0.94} & {\small \textbf{1}} \\ 
& {\small Time complexity (sec)} & {\small \textbf{1.6}} & {\small 34.3} & {\small %
15.2} & {\small 17.5} \\ \hline
& {\small accuracy} & {\small 0.83} & {\small 0.91} & {\small 0.94} & 
{\small \textbf{1}} \\ 
{\small WHITED} & {\small F-score} & {\small 0.83} & {\small 0.9} & {\small %
0.93} & {\small \textbf{1}} \\ 
& {\small Time complexity (sec)} & {\small \textbf{0.071}} & {\small 6.8} & {\small %
3.8} & {\small 3.4} \\ \hline
\end{tabular}

\end{center}
\end{table}

\subsection{Comparison with other classifiers}
Under the classification stage, in addition to the developed BDT classifier, other machine learning (ML) models were also deployed in the experiments including: support vector machine (SVM), deep neural networks (DNN), K-nearest neighbors (KNN), decision tree (DT). These models were proceeded with respect to different classification parameters. Table \ref{classComp} provides the obtained results in terms of the accuracy and F-score. It is shown that the BDT classifiers surpasses other ML algorithms for all the datasets examined under this framework.

\begin{table}[t!]
\caption{Accuracy of the appliance identification system based on FNPA-QR with reference various ML classifier}
\label{classComp}
\begin{center}

\begin{tabular}{l|c|c|c|c|c|c|c}
\hline
{\small ML } & {\small Classifier} & \multicolumn{2}{|c|}{\small ACS-F2} & 
\multicolumn{2}{c|}{\small GREEND} & \multicolumn{2}{c}{\small WHITED} \\ 
\cline{3-8}\cline{3-7}
{\small algo} & {\small \ parameters} & {\small accuracy} & {\small F-score} & {\small %
accuracy} & {\small F-score} & {\small accuracy} & {\small F-score} \\ \hline
{\small SVM} & {\small Linear Kernel} & {\small 0.94} & {\small 0.93} & 
{\small 0.95} & {\small 0.95} & {\small 0.98} & {\small 0.98} \\ \hline
{\small SVM} & {\small Quadratic kernel} & {\small 0.92} & {\small 0.91} & 
{\small 0.93} & {\small 0.93} & {\small 0.9} & {\small 0.86} \\ \hline
{\small SVM} & {\small Gaussian kernel} & {\small 0.94} & {\small 0.93} & 
{\small 0.95} & {\small 0.95} & {\small 0.93} & {\small 0.9} \\ \hline
{\small KNN} & {\small K=1/Euclidean distance} & {\small 0.97} & {\small 0.97%
} & {\small 0.98} & {\small 0.98} & {\small 0.96} & {\small 0.93} \\ \hline
{\small KNN} & {\small K=10/Weighted euclidian dis} & {\small 0.96} & 
{\small 0.95} & {\small 0.98} & {\small 0.98} & {\small 0.95} & {\small 0.92}
\\ \hline
{\small KNN} & {\small K=10/Cosine dist} & {\small 0.94} & {\small 0.94} & 
{\small 0.96} & {\small 0.96} & {\small 0.92} & {\small 0.87} \\ \hline
{\small DT} & {\small Fine, 100 splits} & {\small 0.98} & {\small 0.97} & 
{\small 0.99} & {\small 0.99} & {\small 0.93} & {\small 0.91} \\ \hline
{\small DT} & {\small Medium, 20 splits} & {\small 0.93} & {\small 0.93} & 
{\small 0.96} & {\small 0.93} & {\small 0.94} & {\small 0.9} \\ \hline
{\small DT} & {\small Coarse, 4 splits} & {\small 0.92} & {\small 0.89} & 
{\small 0.94} & {\small 0.92} & {\small 0.91} & {\small 0.88} \\ \hline
{\small DNN} & {\small 50 hidden layers} & {\small 0.96} & {\small 0.95} & 
{\small 0.98} & {\small 0.98} & {\small 0.97} & {\small 0.96} \\ \hline
{\small EBT} & {\small 30 learners, 42 k splits} & {\small \textbf{0.99}} & {\small %
\textbf{0.98}} & {\small \textbf{1}} & {\small \textbf{1}} & {\small \textbf{1}} & {\small \textbf{1}} \\ \hline
\end{tabular}

\end{center}
\end{table}

\section{Conclusion} \label{sec4}

This paper presented a robust non-intrusive appliance identification system using TD descriptors. More specifically, our focus was on the use of the FNPA-QR as a dimensionality reduction module. FNPA-QR played an important role since it i) reduced the amount of data samples in the feature vectors, ii) decreased the distance between data points of the same appliance category and increased the distance among the samples from dissimilar classes. Moreover, a BDT classifier was designed that further enhanced the identification accuracy. Consequently, the results of the evaluation were promising, since they showed that using the proposed TD descriptor based FNPA-QR and BDT classifier, it was possible to reach an optimal identification accuracy and also outperform other dimensionality reduction techniques and ML models.

\section*{Acknowledgements}\label{acknowledgements}
This paper was made possible by National Priorities Research Program (NPRP) grant No. 10-0130-170288 from the Qatar National Research Fund (a member of Qatar Foundation). The statements made herein are solely the responsibility of the authors.

%
%

\end{document}